\documentclass{aa}
\usepackage[dvips]{graphicx}

\usepackage{rotating}
\usepackage{aalongtable,lscape}
\usepackage{amssymb,amsmath}
\usepackage{latexsym}
\usepackage{psfig}
\usepackage{epsfig}
\usepackage{amsbsy}
\newcommand{\Msun} {M$_\odot$}
\newcommand{\Lsun} {L$_\odot$}
\newcommand{\Rsun} {R$_\odot$}
\newcommand{\Teff} {T$_{\rm{eff}}$}
\newcommand{\Tstar} {T$_{\rm{eff}}$}
\newcommand{\Lstar} {L$_\star$}
\newcommand{\Mstar} {M$_\star$}
\newcommand{\Rstar} {R$_\star$}

\newcommand{\Av} {A$_V$}
\newcommand{\um} {$\mu$m}
\newcommand{\AJ} {A$_J$}
\newcommand{\Aj} {A$_J$}

\newcommand{\Myr} {M$_\odot$/yr}
\newcommand{\kms} {km/s}
\newcommand{\Ha} {H$_\alpha$}

\newcommand{\Pab} {Pa$_\beta$}
\newcommand{\Pag} {Pa$_\gamma$}
\newcommand{\Roph} {$\rho$-Oph}

\newcommand{\Brg} {Br$_\gamma$}
\newcommand{\Rl} {Pa$_\beta$/Br$_\gamma$}
\newcommand{\R} {{\it R}}

\newcommand{\Macc} {$\dot M_{acc}$}
\newcommand{\Lacc} {L$_{acc}$}

\newcommand{\ISO} {ISO-Oph~}

\newcommand{\simless}{\mathbin{\lower 3pt\hbox
      {$\rlap{\raise 5pt\hbox{$\char'074$}}\mathchar"7218$}}} 
\newcommand{\simgreat}{\mathbin{\lower 3pt\hbox
     {$\rlap{\raise 5pt\hbox{$\char'076$}}\mathchar"7218$}}} 

\begin{document}

\title{Accretion in  $\rho$ Ophiuchi brown dwarfs: infrared  hydrogen line ratios
\thanks{
Based on observations collected at the European Southern Observatory, Chile.
  Program 075.C-265.}
}

\author{
T. Gatti\inst{1},
L. Testi\inst{1},
A. Natta\inst{1},
S. Randich \inst{1}
\and
J. Muzerolle \inst{2}
}
\institute{
    Osservatorio Astrofisico di Arcetri, INAF, Largo E.Fermi 5,
    I-50125 Firenze, Italy 
\and
 Universit\`a  di Firenze, Dipartimento di Astronomia, Largo E.Fermi 5,
    I-50125 Firenze, Italy
\and
Steward Observatory, University of Arizona
}

\offprints{tomcat@arcetri.astro.it}
\date{Received ...; accepted ...}

\authorrunning{Gatti et al.}
\titlerunning{Accretion in $\rho$-Oph}

\abstract
{Mass accretion rate determinations are fundamental for an understanding
of the evolution of pre-main sequence star circumstellar disks.}
{Magnetospheric accretion models are used to derive values of the
mass accretion rates in objects of very different properties, from
brown dwarfs to intermediate-mass stars;
we  test the validity of these models in the brown dwarf
regime, where the stellar mass and luminosity, as well as the
mass accretion rate, are much lower than in T Tauri stars.}
{We have measured simultaneously two infrared hydrogen lines, \Pab\ and
\Brg, in a sample of 16 objects in the star-forming region \Roph. The sample includes 7 very low mass objects and brown dwarfs and 9 T Tauri stars.}
{ Brown dwarfs where both lines are detected
have a ratio \Pab/\Brg of $\sim 2$. Larger values,
$\simgreat 3.5$, are only found among the T Tauri stars. The low line ratios
in brown dwarfs
indicate that the lines cannot originate in the column of gas accreting from the disk onto the star along the magnetic field lines, and we suggest
that they form instead in the shocked photosphere, heated to temperatures
of $\sim 3500$ K. If so,
in analogy to veiling estimates in T Tauri stars,
the hydrogen infrared line fluxes may provide a  reliable 
measure of the accretion rate in brown dwarfs.}
{}
\maketitle

\section {Introduction}


In  pre-main 
sequence stars,
circumstellar disks feed matter onto their central stars over a period 
of a few
million years.  This accretion process does not significantly alter the 
properties
of the star during most of this time; however, it can have a significant 
influence
on the lifetime and evolution of the disks and of the planetary systems 
that may form.
The physical conditions of the accretion process, in particular the mass 
accretion rate,
can be investigated by studying the line and continuum emission produced by
the accreting material.
The model which has been more successful in explaining
the accretion phenomena in T Tauri stars (TTS) is  magnetospheric
accretion, where  the disk is truncated by the effect of a stellar
magnetic field near or within the corotation radius. The disk material,
which is slowly drifting radially toward the star, reaches the
truncation radius and is then lifted above
the disk midplane and accretes onto the star along  magnetic field
lines, impacting on the star at approximately the escape velocity.
A shock forms on the stellar surface, creating a hot spot which emits
the excess UV continuum and lines observed in TTS.
Optical and IR line emission is expected to come from the accreting
columns of gas, where the temperature has to be of the order of 10000 K.
 Magnetospheric accretion models
have been developed in detail under the assumption of dipole magnetic
field by, e.g.,  Hartmann et al.~(\cite{Hea94}), Calvet and Gullbring 
(\cite{CG98}), Muzerolle et al.~(\cite{Mea98a}, \cite{Mea01}),
Lamzin (\cite{Lam95}, \cite{Lam98}),  to predict the expected
amount of veiling  and line profiles and intensities. They have been
used   to derive
accretion rates for TTS, brown dwarfs (BDs), some intermediate-mass
TTS and Herbig Ae stars 
(Gullbring et al.~\cite{Gea98}, Muzerolle et al.~\cite{Mea01}, \cite{Mea03},
\cite{Mea04}, \cite{Mea05},
Natta et al.~\cite{Nea04}, \cite{Nea06},
Calvet et al.~\cite{Cea04}, Garcia Lopez et al.~\cite{Gea06}).

Testing magnetospheric accretion models is therefore very important and 
extensive checks have been carried out in several of the above papers.
They have been mostly focussed on TTS, where it is possible
to observe the accretion-driven emission over a large range of wavelengths,
both in lines and continuum. Although it is likely
that the topology of the stellar magnetic field is much more complex than
the dipole field for which most models have been developed, 
the  models seem able to account satisfactorily for TTS activity
(see Bouvier et al.~(\cite{Bou_PPV} and references therein).

The situation is different for BDs, where at present the accreting matter 
is usually detected through  optical and
near-IR line emission. In fact, the vast majority of
the existing estimates of 
\Macc\ in BDs are derived by fitting the observed \Ha\ profiles with
emission in magnetospheric accreting columns (Muzerolle et al.~\cite{Mea03},
\cite{Mea05}, Natta et al.~\cite{Nea04}) or by means of secondary
indicators, such as the luminosity of the hydrogen near-IR lines or
of the Ca IR triplet, which have been calibrated using the \Ha-measured
accretion rates (Natta et al.~\cite{Nea04}, Mohanty et al.~\cite{Subu05}).

This paper reports the result of a project aimed at testing the capability
of current magnetospheric models to describe quantitatively
the accretion in   very low mass objects and BDs by
comparing the observed ratios of two IR hydrogen recombination lines,
namely \Pab\ and \Brg, to the model predictions. 
These lines  are seen in emission in most accreting TTS and BDs;
their ratio can be measured relatively easily, and is not
very sensitive to extinction. 
Muzerolle et al.~(\cite{Mea01}) models predict that
for the relativly large accretion rates of TTS ($\simgreat 10^{-8}$ \Myr),
both lines are optically thick and have a ratio \Rl$\sim 4$,
typical of thermalized emission in a gas
at about $10^4$ K. As \Macc\ decreases below  $\sim 10^{-8}$ \Myr,
first \Brg\ and then \Pab\ become optically thin, and their ratio increases
 to  much larger values. The sample of TTS studied
by Muzerolle et al.~(\cite{Mea01}) does not have objects with very low \Macc, and this particular
prediction could not be tested. However, 
in our study of the accretion properties in \Roph\
BDs (Natta et al.~\cite{Nea04}) we found that  two objects, with \Macc$\simless 10^{-9}$ 
\Myr, have
\R $\sim 2$, inconsistent with
magnetospheric accretion models.

Unfortunately, the two lines  had not been
observed simultaneously, and a chance low ratio due to time
variability could not be ruled out.
We present here the results of  simultaneous observations of \Pab\ and \Brg\ in
a sample of 16 objects in \Roph, 7 of which are very low mass objects and BDs and
the other 9 higher mass stars, added for comparison with the Taurus TTS.
The observations and results are presented in Sec.2 and 3, respectively. The
reliability of the derived line ratios is discussed in Sec.4. Comparison with
the magnetospheric accretion models follow in Sec.5. Sec.6 contains  summary and
conclusions.

\section{Observations and data reduction}

\subsection{Sample}
\label{s2_1}

\begin{table*}
\begin{center}
\caption {  Observational and physical parameters of the $\rho$-Oph objects.
Column 1: name of object; 2: spectral type; 3: J magnitude; 4: K magnitude;
5: J-band extinction; 6: luminosity; 7: mass; 8: effective temperature; 9:
\Pab\ equivalent width; 10: \Brg\ equivalent width; 11: \Rl\  flux ratio; 
12: mass accretion rate.} 
\begin{tabular}{lccccccccccc}
\hline 
\multicolumn{1}{c}{Obj.} &
\multicolumn{1}{c}{Spectral} &
\multicolumn{1}{c}{J} &
\multicolumn{1}{c}{K} &
\multicolumn{1}{c}{$A_{j}$} &
\multicolumn{1}{c}{$L_{*}$} &
\multicolumn{1}{c}{$M_{*}$} &
\multicolumn{1}{c}{$T_{eff}$} &
\multicolumn{1}{c}{Ew(Pa$\beta$)} &
\multicolumn{1}{c}{Ew(Br$\gamma$)} &
\multicolumn{1}{c}{R} &
\multicolumn{1}{c}{$\dot{M}$} \\

\multicolumn{1}{c}{} &
\multicolumn{1}{c}{Type} &
\multicolumn{1}{c}{(mag)} &
\multicolumn{1}{c}{(mag)} &
\multicolumn{1}{c}{(mag)} &
\multicolumn{1}{c}{($L_{\odot}$)} &
\multicolumn{1}{c}{($M_{\odot}$)} &
\multicolumn{1}{c}{(K)} &
\multicolumn{1}{c}{(\AA)} &
\multicolumn{1}{c}{(\AA)} &
\multicolumn{1}{c}{} &
\multicolumn{1}{c}{($M_{\odot}/yr$)} \\

\hline \hline

\ISO 002a & M0 & 12.838 & 9.545 & 3.1 & 0.55 & 0.4 & 3850 & 1.9$\pm0.2$ & 2.1$\pm0.3$ & 1.6$\pm0.3$ & -8.73\\
\ISO 023 & M7 & 14.844  & 12.143 & 2.4  & 0.04 & 0.04 & 2650 & 1.2$\pm0.2$ & 1.6$\pm0.2$ & 1.7$\pm0.4$ & -9.69\\
\ISO 030  & M6 & 12.57 & 10.92 & 0.9 & 0.07 & 0.06 & 2700 & $<0.9$ & $<6.0$ & -- &$<$-9.54\\
\ISO 033 & M8.5 & 16.45 & 13.93 & 2.2 & 0.01 & 0.015 & 2400 & $<0.7$ & $<2.7$ & -- &$<$-10.83\\
\ISO 037 & K5 & 15.052 & 10.224  & 3.9 &  0.16 & 0.7 & 4350 & 4.4$\pm0.5$ & 0.8$\pm0.3$ &  3.8$\pm1.5$ & -9.61\\
\ISO 083 & K9 & 12.257 & 9.251 & 3.0 & 0.81 & 0.4 & 3900 &  2.1$\pm0.2$ & $<0.9$ &  $>5.2$ & -8.34\\
\ISO 102 & M6 & 12.433 & 10.766 & 0.9 & 0.08 & 0.06 & 2700 & 1.4$\pm0.3$ & 1.8$\pm0.3$ & 2.0$\pm0.5$ & -9.17\\
\ISO 105 & K9 & 12.547  & 8.915 & 4.0 & 1.5 & 0.35 & 3900 & 1.3$\pm0.2$ & $<3.0$ &  $>1.0$ & -8.03\\
\ISO 115 & M0 & 15.616  & 11.486 & 3.7 &  0.07 & 0.6 & 3850 & 1.1$\pm0.4$ & $<4.8$ & $>0.3$ & -10.81\\
\ISO 117 & K8 & 13.325  & 9.978 & 3.1 &  0.35 & 0.6 & 3950 & 6.5$\pm0.3$ & 3.3$\pm0.1$ & 3.0$\pm0.2$ & -8.63\\
\ISO 155 & K3 & 11.322  & 7.806 & 3.2 & 3.1 & 1.0 & 4730 & 1.3$\pm0.1$ & $<1.8$  & $>0.9$ & -8.28\\
\ISO 160 & M6 & 14.148  & 11.947 & 1.9 &  0.04 & 0.05 & 2700 & 1.2$\pm0.2$ & 2.3$\pm0.4$  & 1.5$\pm0.3$ & -9.65\\
\ISO 163 & K5 & 11.378 & 8.271 &  2.7 &  1.6 & 0.6 & 4350 & 3.3$\pm0.2$ & 1.1$\pm0.4$  & 4.6$\pm1.7$ & -7.89\\
\ISO 164  &  M6 & 13.27 & 11.08 & 1.9 & 0.09  & 0.06 & 2700 & $<0.3$ & $<2.8$ & -- & $<$-9.39\\ 
\ISO 166 & K5 & 10.754 & 8.464 & 2.2 &  1.9 & 0.6 & 4350 & 1.7$\pm0.2$ & $<1.2$  & $>3.9$ & -8.19\\
\ISO 193 & M6 & 13.611 & 11.086 & 2.3 & 0.1 & 0.06 &2650  & 1.7$\pm0.2$ & 1.9$\pm0.4$ & 2.2$\pm0.5$ & -8.88\\

\hline
\end{tabular}
\end{center}
\end{table*}

We have selected 16 objects in \Roph\ which have been previously
studied by Natta et al.~(\cite{Nea02}, \cite{Nea04}, \cite{Nea06});  15 of them
are known
to have \Pab\ in emission, while in one case (\ISO 033) \Pab\ was not detected, and we added it to our sample because it is the lowest mass object
in \Roph\ with a measured J-band spectrum.

The adopted  spectral type, effective temperature, luminosity and mass of
the sample are given
in Table ~1. The spectral classification of objects in
\Roph\ is often very uncertain, as discussed, e.g., by
Luhman and Rieke (\cite{LR99}), Wilking et al.~(\cite{Wea05}),
Doppmann et al.~(\cite{Dea05}), Natta et al.~(\cite{Nea06}).
For   7 very low mass objects and BDs (\ISO 023, \ISO 030, \ISO 033,
\ISO 102, \ISO 160, \ISO 164 and \ISO 193) we adopt the stellar
parameters derived by
Natta et al.~(\cite{Nea02})  by comparing
low resolution simultaneous J, H and K spectra
to template field stars and to model atmospheres. 
The uncertainties are of  order $\pm 150$ K in \Tstar, $\pm 0.2$ dex in
\Lstar, $\pm 1$ mag on \Av.

For the other stars, 
we  estimate the spectral type from the J-band medium resolution
spectra obtained in this paper, as described in the
Appendix. The corresponding \Tstar\ is from Kenyon \& Hartmann (\cite{KH95}).
The extinction \AJ\ is derived from the 2MASS (J-H)
and (H-K)  colors, by de-reddening the objects to
the locus of the classical T Tauri stars  defined by 
Meyer et al.~(\cite{Mea97}), as discussed in Natta et al.~(\cite{Nea06}).
We adopt an extinction law  characterized by E(J-H)/E(H-K)=1.54
(Cardelli et al.~\cite{Cea89}, Kenyon et al.~(\cite{Kea98}),
appropriate for Ophiuchus.
The stellar luminosity is  computed from the de-reddened
J magnitude  and the bolometric correction
corresponding to the assigned spectral type (for a distance $D=150$ pc). 
The uncertainties on \Tstar\ and  \Lstar\ are often large, ranging
from $\pm$100 K to $\pm$650 K in \Teff\ and between $\pm$0.2 and
$\pm$0.5 dex in \Lstar.

\begin{figure}[ht!]
	\begin{center}
 	 \psfig{figure=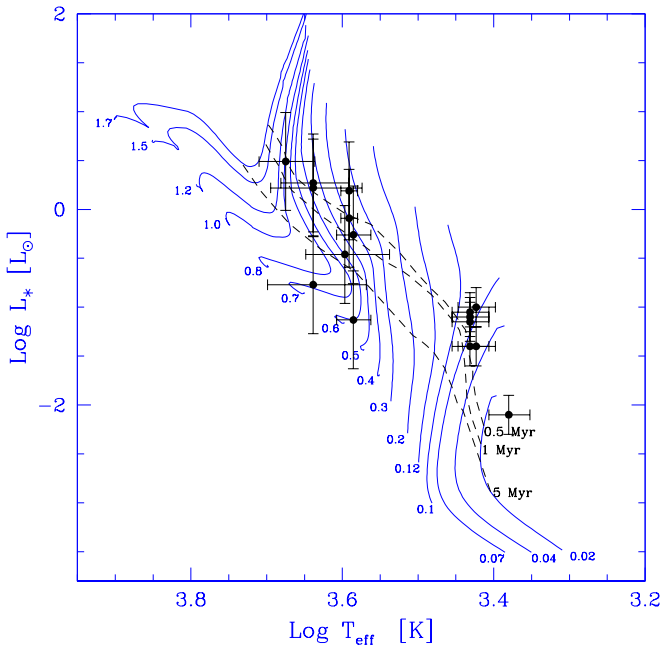, width=9cm,angle=0} 
 \caption {Location of the sample objects on the HR diagram.
The evolutionary tracks (solid lines, labelled with the stellar mass)
are from D'Antona and Mazzitelli (\cite{DM97});
dashed lines show isocrones at 0.5, 1 and 5 Myr, as marked.}
\end{center}
\label{fig-HR}
\end{figure}

The  location of the sample objects
on the HR diagram
is shown in Fig.~\ref{fig-HR}, together with the evolutionary tracks
of D'Antona \& Mazzitelli (\cite{DM97}), used to estimate the object masses.
The sample contains 7 objects with mass $\simless 0.1$ \Msun, and 9 with
\Mstar\ in the range  $\sim 0.35- 1$ \Msun.

\subsection{Near-infrared spectroscopy}

Near-infrared spectra in the J and K bands were obtained for all our targets using the ISAAC near-infrared camera and spectrograph at the ESO-VLT UT1 telescope. The observations  were carried out in Visitor Mode on May 19-20, 2005. We used the 0.6 arcsec slit and the medium resolution grism that offered a $\sim$860 resolution across the wavelength 1.0-1.3 $\mu$m and $\sim$750 at 1.95-2.45 $\mu$m. The observing sequence was composed of a set of pairs of spectra with the target in different positions along the slit, to allow for an efficient and accurate sky subtraction. The on-source times varied from 30 to 60 min depending on target brightness and observing band (the J band observations were typically 30-40$\%$ shorter than the K band ones). Standard calibrations (flats and lamps) and telluric standards spectra were obtained for each observation. Wavelength calibration was performed using the lamp observations.

The \Pab\ and \Brg\ lines were observed as ``simultaneously'' as possible, to minimize the uncertainties due to the expected variability; to maintain a good temporal coherence of the observations of the two emission lines, every target was observed in a time sequence \Brg--\Pab--\Brg.
A complete observing cycle lasted about 2 hours.

The spectra were reduced using standard procedures in IRAF.
After checking that the two \Brg\ data sets were not significantly
different (see \S 4.1), we combined all the exposures to obtain the final spectrum.
The portion of the resulting spectra centered around the \Pab\ and \Brg\ 
lines is shown in Fig.~\ref{fig-spectra}.

\begin{figure}[ht!]
	\begin{center}
 	 \psfig{figure=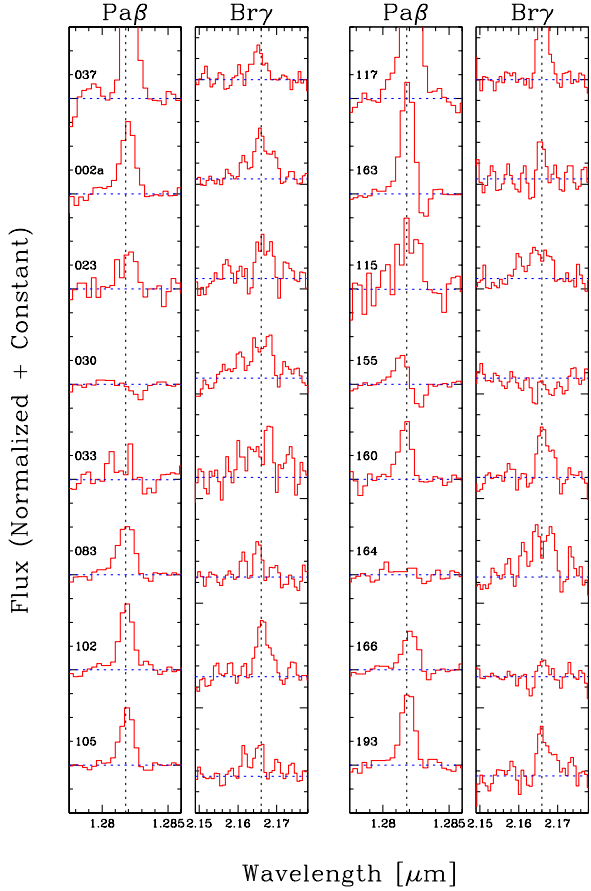, width=9cm,angle=0} 
\linespread{1.0}
	\caption{  Pa$\beta$ (left panel) and Br$\gamma$ (right panel)
spectra of the target objects. The observed fluxes have been normalized to the continuum
near the line and shifted for an easier display.}
	\end{center}
\end{figure}
\label{fig-spectra}

\section{Results}

The  equivalent widths of \Pab\ and \Brg\ were determined by fitting
gaussian profiles to the spectra using standard {\it iraf} routines. 
The uncertainties were estimated 
from the uncertainties on the continuum location and from the
equivalent width of noise spikes. We have checked that the
equivalent width obtained by integrating the observed flux in excess of the continuum in a fixed wavelength range does not deviate from the 
values of the gaussian fit by more than 1$\sigma$.

In 13 objects \Pab\ is clearly seen in emission, with equivalent width between
$\sim 1$ and 6.5 \AA. In 3 cases, the line is not seen and we report
in Table~1 3$\sigma$ upper limits.
In all objects, the emission lines are not resolved, with the
exception of  two of the sample TTS
(\ISO 163 and \ISO 155), where  we clearly detect red-shifted absorption.
However, the resolution of our spectra is not suited for
a study of the line profiles.
\Brg\ is generally weaker, and we have detected it in 8 objects only; in some cases,
the  equivalent widths have large uncertainties.

A comparison of the equivalent width of this paper with the
measurements reported by Natta et al.~(\cite{Nea04}, \cite{Nea06})
shows the importance of measuring line ratios from
spectra obtained simultaneously, as variations of factor of $\sim 2$
are possible. Of the three objects with non detected \Pab, in two cases the current
upper limits are consistent with our previous observations. In one
case (\ISO 164), however,
\Pab\ was clearly detected by Natta et al.~(\cite{Nea04}), with an equivalent width of 0.8\AA,
significantly larger than our current upper limit (0.3 \AA).

We compute line fluxes from the equivalent widths and the flux of the nearby continuum.
We do not  correct the equivalent width
for underlying photospheric absorption,
since the expected equivalent widths are small ($\simless 0.5$ \AA, Wallace
ate al.~\cite{Wea00}) for objects with \Tstar$\simless 5000$ K.
As it was not possible to flux-calibrate the spectra,
we  estimated the continuum flux  using
the  2MASS values of the J and K magnitudes, proceeding as follows. 
The spectra were integrated over the  2MASS filter profiles and scaled
to the measured 2MASS values.
The resulting line fluxes were then corrected for
extinction, according to the extinction law of
Cardelli et al.~\cite{Cea89} for $R_V$=4.2 and the values of \AJ\ given in Table~1.

The  ratio of the line fluxes Pa$_\beta$/Br$_\gamma$ (\R\ in the following)
is  given in Table~1,  
which gives also  the 1$\sigma$
errors on \R, computed
taking into account only the errors on the measured equivalent widths
and on the 2MASS photometry. They range from 3 to $\sim 40$\%, in objects
with very weak \Brg. The possible systematic errors introduced by our 
flux calibration method and the extinction correction are discussed below.

Table~1, last column, gives  the mass accretion rate
\Macc,   computed  from the luminosity of \Pab, using the
relation between L(\Pab) and the accretion luminosity derived by
Muzerolle et al.~(\cite{Mea98b}) and Natta et al.~(\cite{Nea02}):

\begin{equation}
\log{L_{acc}/L_\odot} = 1.36\, \log{L(P_\beta)/L_\odot} + 4
\end {equation}
The mass accretion rate is then computed from \Lacc:
\begin{equation}
\dot M_{acc}=L_{acc} \, R_\star/(GM_\star)
\end {equation}

The uncertainties on \Macc\ are dominated in most cases by uncertainties on the
stellar parameters, and by the scatter of eq.(1) for very low luminosity
objects. They can easily be of the order of one dex  in Log \Macc.

\section {Discussion}

\begin{figure}[ht!]
 	 \psfig{figure=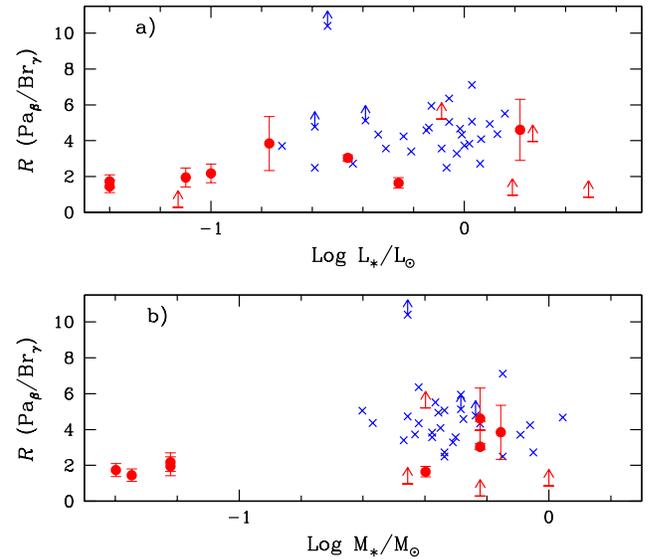,width=9cm} 
 \caption {\R\ is shown as a function of \Lstar\ (Panel a) and of \Mstar\
(Panel b) for the  \Roph\ objects; dots are measured values, arrows lower limits.
Crosses plots \R\ values for the Taurus TTS from Muzerolle et al.~(\cite{Mea98b}).
}
\label{R-stars}
\end{figure}

The values  of \R\  span a rather large
interval,  varying from
less than $2$ to $\simgreat 5$. 
We plot the values of \R\ in Fig.~\ref{R-stars} as a function of the
luminosity of the central object and  of its mass.
Inspection of Fig.~\ref{R-stars} shows that the 4 low luminosity (\Lstar$<0.1$\Lsun),
low mass (\Mstar $\le 0.06$\Msun) objects with measured  \R\ 
(\ISO 023, \ISO 102, \ISO 160, \ISO 193) have
\R$\simless 2$.
Values of \R$\simgreat 3.5$ are only found in more massive objects,
with \Mstar$\ge 0.4$ \Msun.

In fig.~\ref{R-stars}, we have added to our objects the TTS in Taurus for which
the ratio \R\ has been measured by
Muzerolle et al.~(\cite{Mea98b}). For better consistency,
we have recomputed the \R\ values from the line equivalent widths
and \Aj\ 
given by Muzerolle et al.~(\cite{Mea98b}),  and 2MASS photometry.
The differences with the values reported by Muzerolle et al.~(\cite{Mea98b})
are generally small.

The Taurus sample does not include any very low mass object
and is limited to stars with \Mstar $\simgreat 0.25$ \Msun,
but it complements nicely our \Roph\ data for TTS.
In particular, one can see how,
on one side, large values of \R ($\simgreat 3.5$) are observed only in higher mass objects,
while  on the other all the observed  BDs
have \R$\simless 2$.

There are few higher mass objects with \R$\sim2 $. One is \ISO 002,
which will be discussed in Sec.5. For the three Taurus objects
with similarly low \R\ (GI~Tau, FS~Tau and DG~Tau), we note
that the Taurus observations of \Pab\ and \Brg\ are not simultaneous,
and line variability may affect by large factors the \R\ value
of individual objects.
For example, the \Pab\ equivalent width measured for GI~Tau 
at two different epochs by Folha and Emerson (\cite{FE01})
varies by more than a factor of 2.

The low values of \R\  are particularly interesting, as we will
dicuss in the following. However, before doing that,  
we need to  examine the potential sources of error in the
estimates of \R, which may undermine the significance of our
results.

\subsection {Time variability}

Line and continuum variability are well known in pre-main sequence stars of all
masses. We have measured  the two lines simultaneously, using an observational
scheme \Brg\--\Pab\--\Brg\
intended to minimize the effect of variability on \R.
In fact, it turns out that on the time scale of
1--2 hours, spanned by our observations, variability is not large,
and the \Brg\ equivalent widths measured before and after \Pab\ are the same 
within the errors.

\begin{figure}[ht!]
 	 \psfig{figure=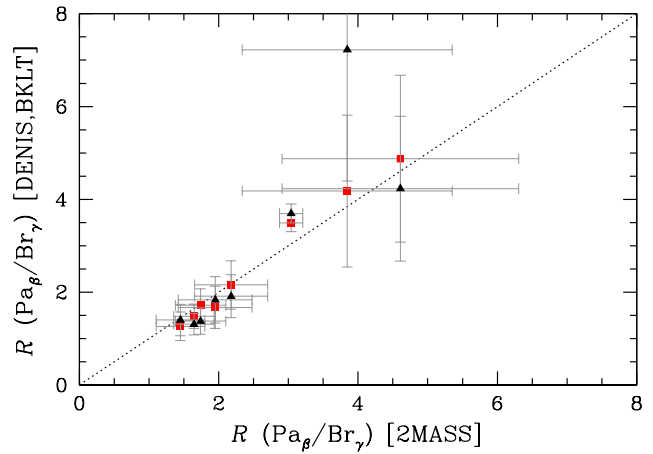,width=9cm} 
\caption {Comparison of \R\  using three different sets
of near-IR photometry to derive the continuum flux at the line
wavelengths. The horizontal axis shows  \R\ computed from the 2MASS
photometry; on the vertical axis, the results obtained using 
DENIS  (squares) and Barsony et al.(\cite{Bea97}) observations (triangles).}
\label{R-var}
\end{figure}

Note, however, that it has not been possible
to  calibrate our spectra photometrically, so that, in order
to compute the line ratio \R,  we had to rely
on  magnitudes from
the literature. 2MASS observations have been performed simultaneously in the three bands J,H,K, so that the values of \R\ assigned to each object can be considered a correct, albeit ``snapshot", value, 
as long as there are
no strong variations in the (J-K) colors of the stars. 
This seems to be indeed the case, as shown by Fig.~\ref{R-var},
which compares the values of \R\ obtained by using the J and K
magnitudes measured by 
DENIS and by Barsony et al.~(\cite{Bea97}), respectively, to
those in Table~1, computed with 2MASS photometry.
All sets of photometric observations have been obtained simultaneously in the
J and K bands.
The differences are always within 2$\sigma$; the five
values of \R\  lower than $\sim 2$ remain low, while the
\R\ values of the order of 4 or larger remain so.

\subsection {The extinction correction}

The values of \R\  depend on the  differential extinction between
the J and K band.
This extinction correction can be 
large  in \Roph\ objects, which are often very embedded, and introduces
an additional uncertainty on \R.
In particular, underestimating  \AJ\ will result in  values of \R\
lower than the true ones, since
\R\ is proportional to $10^{0.22 A_J}$ (for the adopted extinction law).
Therefore, to  push
observed values of \R\ from $\sim 2$ to  $\sim 4$, one needs to
increase the J band extinction by 1.37 mag. 

Such an increase of \Aj\ with respect to the values given in Table 1
is not consistent with the J, H, K spectra and photometry
of the 4 low mass, low \R\ 
objects (\ISO 023, \ISO 102, \ISO 160 and \ISO 193)
studied by Natta et al.~(\cite{Nea02}), who
estimate uncertainties in \Aj\ of $\sim$0.3 mag. 
In fact,
independent estimates of \Av\ for two
objects (\ISO 030
and \ISO 102) by Wilking et al.~(\cite{Wea05}) result in values of
\Aj\ even lower than those adopted in this paper.

It is more difficult to estimate the \Aj\ uncertainties for the 8
stars in our
sample for which \Aj\ is derived from the (J-H)-(H-K) colors.
Luhman and Rieke (\cite{LR99}), using
the Barsony et al.~(\cite{Bea97}) photometry and a more
complex scheme than the simpler one we have implemented,
derive values of \Aj\ for the
5 objects we have in common within 0.3 mag of our estimates.
In one case (\ISO 166) there is also in the literature an
estimate of the extinction from optical spectra (Wilking et al.~\cite{Wea05}), which agrees with our value within 0.1 mag.

 
In summary, although it is difficult to rule out the possibility of a
large error on \Aj\ in some individual object, we think it is unlikely that this
can systematically affect our results. In particular, it does not seem
possible to interpret all the low \R\ values as due to an underestimate
of the extinction.





\section {Comparison with magnetospheric accretion models}

The main aim of this work was to extend toward very low mass
objects the
comparison of  the observed ratios \Rl\
to the prediction of magnetospheric accretion models.
In these models,  gas is channeled from the disk onto the star
along the lines of the stellar magnetic field and a shock forms
where the  accreting matter impacts on the stellar photosphere.
The accretion luminosity is emitted mostly by the heated
photosphere below the shock (the hot spots on the stellar surface),
with the rest  coming out from the pre-shock region.
In accreting TTS, the heated photosphere and the pre-shock
region are responsible for the  continuum
which veils the optical spectrum and for the strong ultraviolet continuum
and line  spectrum seen in these stars. The optical lines
are believed to form not in the shocked region, which is
very optically thick, but in the accreting columns, heated to roughly
uniform temperatures (of order 8000 - 12000 K) not by the shock
emission but by some alternative mechanism, possibly wave heating.
These models have been very successful in explaining the complex
phenomenon of TTS activity (e.g., Bouvier et al.~\cite{Bou_PPV} and references therein).


In the magnetospheric accretion models, for fixed stellar 
parameters and
field geometry, R depends on the mass accretion rate Macc and temperature
of the accreting gas.  As mentioned, the gas temperature
is not computed self-consistently, but has been constrained by fitting 
different lines
of various atomic species  (see Muzerolle et al. \cite{Mea01}).
Within these constraints, the infrared lines are thermalized for the gas
densities expected for average values of \Macc.  Both the line
luminosities and \R\ do not change with decreasing density until \Macc $\simless 10^{-8}$ \Msun/yr, at which point the line optical depth goes below unity first at 
\Brg\
and then \Pab\, and R quickly increases to very large values.
There are additional free parameters, namely the exact location of
the base of the accreting region and its inclination to the line of sight,
which, however, do not change this basic picture.

\begin{figure}[ht!]
\begin{center}
  \psfig{figure=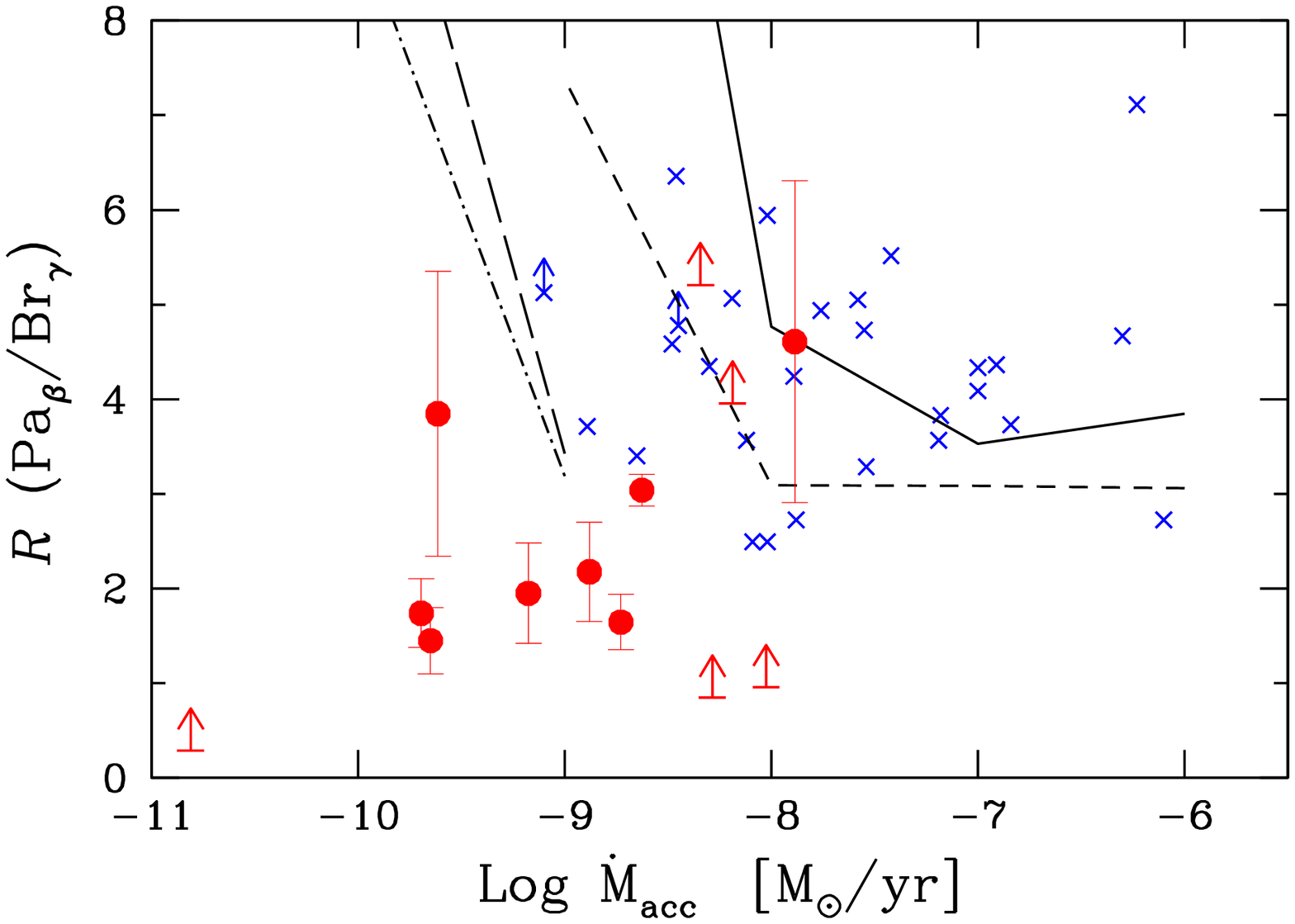, width=9cm,angle=0} 
\linespread{1.0}
\caption{ \R\ as a function of \Macc\ for the \Roph\ objects;
dots show actual measurements, arrow lower limits. \Macc\
has been derived from the \Pab\ luminosity, as described in the
text. Crosses show the location of Taurus TTS  (from 
Muzzerolle et al.~\cite{Mea98b}). The different lines show the predictions 
of magnetospheric accretion models which roughly fit the UV veiling 
and the \Ha\ profiles. 
Solid and short-dashed lines show models for TTS (\Mstar=0.5 \Msun,
\Rstar=2 \Rsun) with inner and outer magnetosphere
boundaries at the disk
of 2.2--3 \Rstar (solid) and 2.8--3 \Rstar (short-dashed).
The temperature of the accreting gas increase from 7000 K 
(for \Macc = $10^{-6}$ \Myr) to 12000 K for \Macc=$10^{-9}$ \Myr (see Muzerolle et al.~\cite{Mea01} for details).
The other two lines show models for typical BD parameters (\Mstar=0.05 \Msun,
\Rstar=0.5 \Rsun), magnetosphere boundaries 2.2--3 \Rstar, temperature
of 12000 K (long-dashed) and 10000 K (dot-dashed), respectively.
}
\label{R-macc}
\end{center}
\end{figure}

The predictions of magnetospheric accretion models are compared
to our results in Fig.~\ref{R-macc}, which
shows that the models
reproduce quite well the bulk of the \R\ values in the Taurus TTS
(Muzerolle et al.~\cite{Mea01}).

The situation is different for the \Roph\ BDs, which have
much lower \Macc\ than TTS
(e.g., Muzerolle et al.~\cite{Mea03}, \cite{Mea05}; 
Natta et al.~\cite{Nea04}; Mohanty et al.~\cite{Subu05}).
Fig.~\ref{R-macc} shows model prediction for stellar parameters typical
of BDs and \Macc$\leq 10^{-9}$ \Myr.
The gas temperature required to fit intensity and profiles of \Ha\
is T$\simgreat 8000$ K.
In these models \R\ $\sim 3.5$ for \Macc $=10^{-9}$ \Myr, and increases
rapidly with decreasing \Macc. 
It is clear that they cannot reproduce the observed low
\R\ values of the \Roph\ BDs.

\begin{figure}[ht!]
\begin{center}
  \psfig{figure=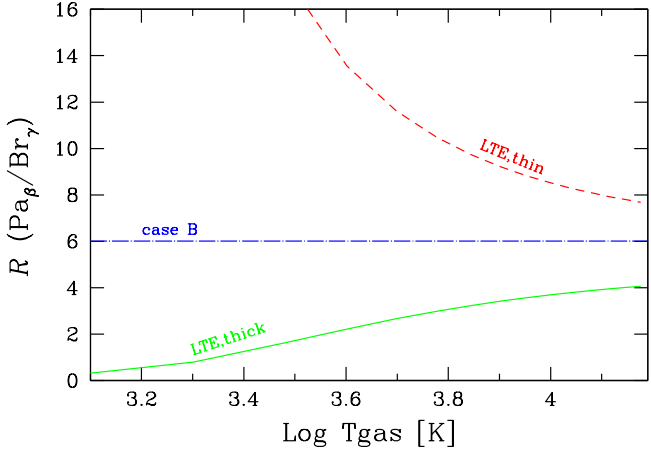, width=9cm,angle=0} 
\linespread{1.0}
\caption{\R\ is plotted as a function of the gas temperature T$_{gas}$ for
three textbook cases. The solid line shows the case of optically thick lines
with LTE level population; the dashed line the case of LTE, optically thin
lines. The dash-dotted line the value of \R\ for case B recombination.}
\label{R-theory}
\end{center}
\end {figure}

\R$\sim 2$ is obtained if the lines form in
a cold,  optically thick gas. This is shown in
Fig.~\ref{R-theory}, which plots  the \R\ values expected as a function of
the gas temperature $T$ in different physical situations. 
The  line at \R$\sim 6$ shows the
predictions of Case B, i.e., the value of \R\ when the lines are
optically thin and the level population is dominated by radiative cascade
from the continuum.
If the hydrogen levels from which the near-IR lines originate
($n \ge 3$) are in LTE, the dependence of
\R\ on the temperature is different for optically thin
and optically thick lines. When the  lines are optically thin, \R$\simgreat 8$
for temperature $T\sim 10000$K and increases for decreasing $T$ (dashed line).
On the contrary, when the lines are optically thick, 
\R\ decreases for decreasing temperature,
from  $\sim 4$
for $T\simgreat$ 8000 K to $\sim 2$ for $T\sim 3500$ K. 

\begin{figure}[ht!]
\begin{center}
  \psfig{figure=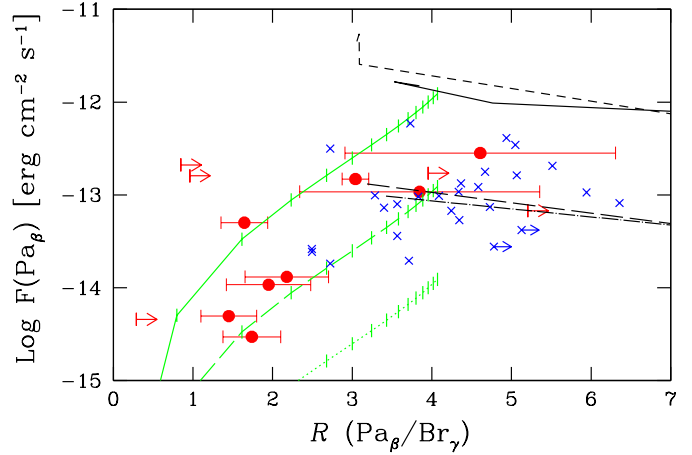, width=9cm,angle=0} 
\linespread{1.0}
\caption{\Pab\ flux versus \R\ for Ophiuchus (dots and arrows) and Taurus
(crosses). The three curves ending at \R$\sim 4$ show the 
LTE, optically thick line emission case. Along each line, the temperature
increases from 1000 K to 15000 K, in steps of 1000 K
(tick marks);  the first mark corresponds to T=2000 K on the top curve, to
T=3000 K for the middle curve and to T=5000 K for the lowest one.
Note that the values of \R\ depend on T only.
Each curve corresponds to a different value of the emitting area,
$2\times 10^{22}$ cm$^2$ (solid line), 
$2\times 10^{21}$ cm$^2$ (long-dashed line) and $2\times 10^{20}$ cm$^2$ 
(dotted line) from top to bottom, respectively. In all
cases, we have assumed a \Pab\ FWHM of 100 km s$^{-1}$;
the \Pab\ flux is proportional to the emitting area times the line width.
Magnetospheric accretion models (same as in Fig.~\ref{R-macc})
are shown by the roughly horizontal lines (same as Fig.~\ref{R-macc}). 
Slightly lower values of the gas temperature in the accreting columns
will give lower value of F(\Pab), while still roughly fitting the
observations. Note, however,  that in no case \R$<3$. 
}
\label{fpab-R}
\end{center}
\end {figure}

If  indeed the lines are thermalized,  the size of the 
emitting region can be estimated from Fig.~\ref{fpab-R},
which plots  the luminosity of \Pab\  as a function of \R\ for the
Ophiuchus and Taurus sample.
The three curves show 
the location of thermalized, optically thick
IR line emission for different values of the projected area of
the emitting region, from $2\times 10^{22}$ to $2\times 10^{20}$ cm$^2$ 
and a FWHM line width of 100 km s$^{-1}$. 
The four BDs (\ISO 23, \ISO 102, \ISO 160 and \ISO 193)
with  \R$\sim 2$,
are well fitted by black body emission at T$\sim 3500$ K and
area $\sim 2\times 10^{21}$ cm$^2$. This is a small
fraction of the projected photospheric surface, varying between
$\sim 6$ and 15\% in the four objects.
Note that at the resolution of our data ($\sim 350$ \kms)
the BDs lines are not resolved; as the \Pab\ flux is proportional to
the product of the emitting area times the line width, larger or
smaller values of the latter will result in a proportionally 
smaller or larger values of the emitting surface.

These properties (i.e., temperature higher than \Tstar,
small covering fraction, high optical depth)
remind those of the hot spots caused by magnetospheric accretion
on the stellar surface of classical TTS.

\begin{figure}[ht!]
\begin{center}
  \psfig{figure=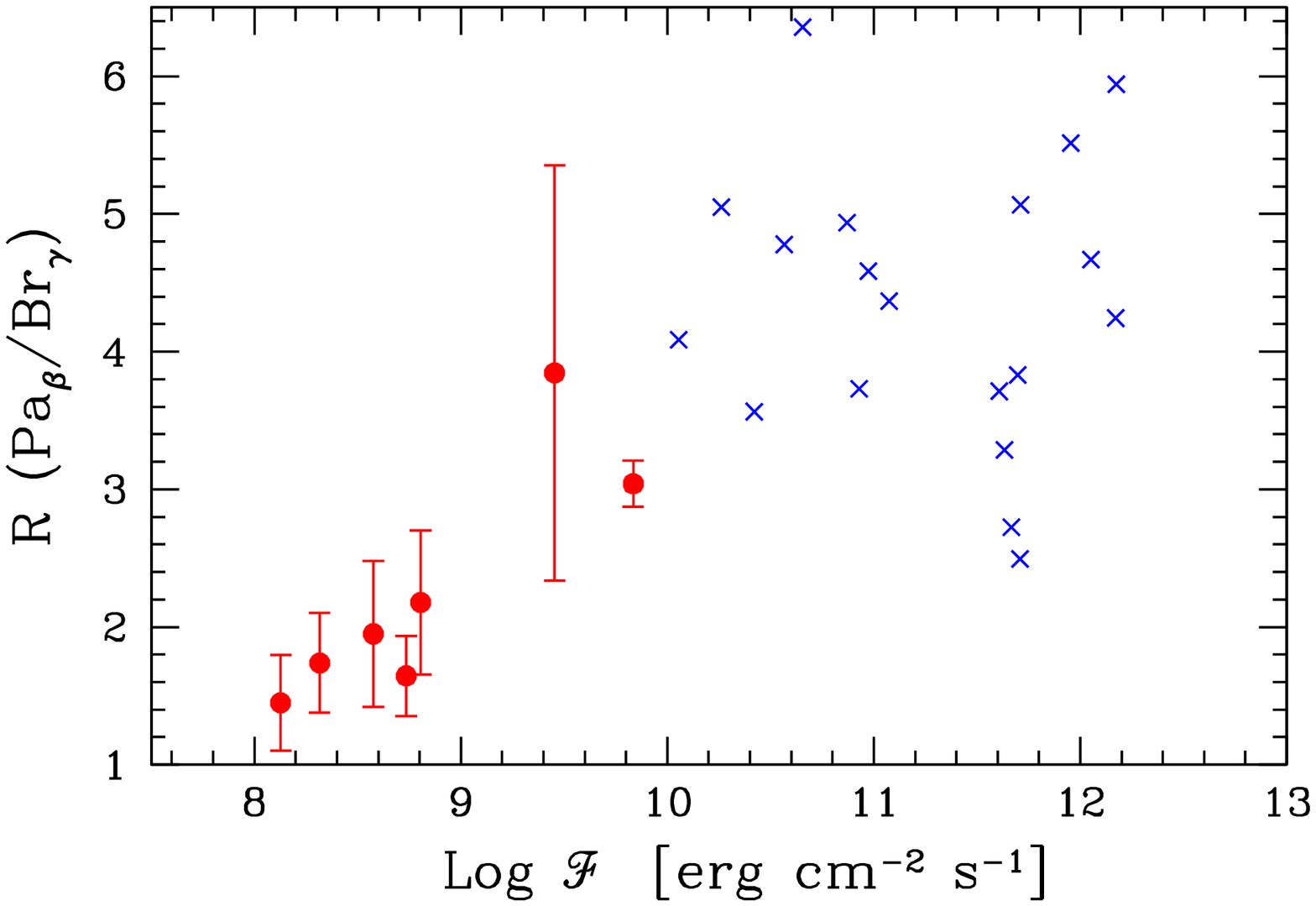, width=9cm,angle=0} 
\linespread{1.0}
\caption{\R\ values  plotted against the energy flux of the accretion flow
$\cal F$. Estimates for the \Roph\ objects (dots) have been derived from 
their measured \Macc, stellar parameters and an emitting area estimated by
tring heir location on the F(\Pab)--\R\ with the predictions
of LTE, optically thick emission (Fig.~\ref{fpab-R});
two objects (\ISO 037 and \ISO 163) with \R$\simgreat 4.5$ no estimate of $\cal F$ was possible.
The values for Taurus TTS have been computed from \Macc\
and stellar parameters given by Muzerolle et al.~(\cite{Mea98b}) and the
fraction fo stellar surface covered by the accretion spots computed
by Calvet and Gullbring (\cite{CG98}). In all cases we have used eq.(11)
of Calvet and Gullbring (\cite{CG98}).
}
\label{accflux}
\end{center}
\end {figure}

Models of the  accretion shock  have been computed
by Calvet and Gullbring (\cite{CG98}) for
the parameters typical of TTS. They showed that the emission 
originates from three different regions, the pre-shock accreting columns,
the post-shock gas and the heated photosphere below the shock (the
hot spot). The physical properties of these regions are controlled
by the energy flux of the accretion flow $\cal F$:

\begin{eqnarray}
{\cal F} = 9.8 \times 10^{10}\> {\rm erg\, cm^{-2}\, s^{-1}} \>
\Big({\rm {{\dot M_{acc}}\over{10^{-8}\, M_\odot\, yr^{-1}}}}\Big) \nonumber \\
\Big({\rm {{M_\star}\over{0.5 M_\odot}}}\Big) \>\> \Big({\rm {{R_\star}\over{2R_\odot}}}\Big)^{-3} \>\>
\Big({{f}\over{0.01}}\Big)^{-1}
\end{eqnarray}
where $f$ is the fraction of the stellar surface covered by the accretion spots.
In TTS $\cal F$ ranges between $3 \times 10^{10}$ and $10^{12}$ erg cm$^{-2}$ s$^{-1}$; with these values of $\cal F$, the heated photosphere is very
optically thick, and emits a continuum spectrum with temperatures
much higher than \Tstar, which accounts for the UV veiling.
The models are quite complex and cannot be simply
scaled to the BD parameters. However, we note that the energy flux of
the accretion flow in the four \Roph\ BDs is ${\cal F}\sim 5\times 10^7 - 5\times 10^8$ erg cm$^{-2}$ s$^{-1}$,
much lower than in TTS (see Fig.~\ref{accflux}).  
From the Calvet and Gullbring models,
one expects that both the
enhancement of the temperature
with respect to the undisturbed photosphere,
and the depth of the affected region decreases with ${\cal F}$, so that
it seems 
possible to obtain  line emission with the required 
properties from the heated photosphere. 

Fig.~\ref{accflux} plots the observed values of \R\ as a function of
${\cal F}$ for Taurus and \Roph\ objects. For Taurus, we have computed
${\cal F}$ from the accretion rates, stellar parameters and  the
surface coverage of the accretion columns given by Calvet and
Gullbring (\cite{CG98}), using their eq.(11). For the \Roph\ objects, we 
take  the temperature and the area of the emitting region from the
location of the object in Fig~\ref{fpab-R}, assuming black body emission.
This is possible for the 6 objects with \R$\simless 3.5$.
One can see that 5 of them (the 4 BDs and \ISO 002), which have \R$\sim 2$, 
also have ${\cal F}$ much smaller than the Taurus TTS.
The sixth object (\ISO 117) has \R$\sim 3.5$ and ${\cal F}$ similar to
the lowest values of Gullbring \& Calvet.
The correlation between \R\ and ${\cal F}$ stops at ${\cal F}\sim 10^{10}$;
for higher ${\cal F}$, all the objects have similar values of \R.

We propose a scenario  where for high \Macc\ the optical and infrared
continuum emission of the 
heated photosphere is optically thick and all the
hydrogen lines (optical and IR) observed in emission
originate from  the  accreting columns of gas.
When \Macc\ decreases below a threshold value the continuum
emission of the hot spots becomes optically thin
and the lines appear in emission; at the same time,
the  density of the accreting gas decreases and its line
spectrum gets weaker. As a consequence,
lines of lower optical depth  have
an increasing contribution from the shocked regions.
In BDs, this region dominates the emission of the near-IR hydrogen
lines, while optical lines, such as \Ha\, still originates mostly
in the accreting columns of gas.
If this will be confirmed by further observations and by model
calculations,
the IR hydrogen lines may provide a
measurement of \Macc\ in BDs in analogy to continuum veiling in TTS,
less model-dependent than the \Ha\ profiles used so far.

A similar picture is suggested
by Edwards et al.~(\cite{Eea06}), who   obtained
high resolution profiles of \Pag\ in classical Taurus TTS.
They find that the \Pag\ 
width correlates with the  veiling at 1 \um,
which may be a tracer of \Macc.
The width of lines forming in the accreting gas columns should 
not depend on \Macc, for fixed stellar and disk parameters,
and Edwards et al. argue that their results can be understood
if the relative contribution  of the
accretion shock  to the \Pag\ emission increases 
as \Macc\  decreases.

\section {Summary and conclusions}

This paper presents the results of simultaneous measurements of \Pab\
and \Brg\ in a sample of 16 objects in the \Roph\ region.
We measure the ratio of the line fluxes \R=\Rl\ in 8 objects, 4 of which
are spectrally confirmed very low mass stars or brown dwarfs (BDs in this
paper). In
5 other cases we could not detect \Brg, and we estimate lower limits
to \R.

The 4  BDs all have \R$\sim 2$; of the other stars, one has \R$\sim 2$, while
the others have \R$\ge 3.5$, similar to the values measured in accreting TTS
in Taurus (Muzerolle et al.~\cite{Mea98b}). 
We  discuss the measured \R\ values in the context of magnetospheric accretion models where
hydrogen recombination lines form in the accreting columns of gas that connect the disk to the star.
The gas is heated to temperatures of $\sim 8000 - 12000$K, and the predicted
values of \R\ are in all cases $\ge 3.5$.
The low \R\ observed in the \Roph\ BDs cannot be explained by
these models.

We propose that, at the low \Macc\ typical of the BDs,
the hydrogen near-IR lines form instead in the shock-heated stellar
photosphere,  and
discuss how it is possible
to produce in these regions the required low temperature and high
optical depths in the line (but not in the continuum)
that will result \Pab\ and \Brg\ emission with in  \R$\sim 2$.
This possibility needs to be investigated with the help of models
similar to those of Calvet and Gullbring (\cite{CG98}),
which compute the emission of the various components of magnetospheric
accretion (accreting column, shock-heated photosphere and post-shock gas) 
for low \Macc\ values and the parameters typical of BDs.

Finally, we note that
the luminosity of the hydrogen IR recombination lines has been extensively used to
derive the accretion rate in large sample of objects, especially
when UV veiling could not  be measured (e.g., Natta et al.~\cite{Nea06}).
The calibration of the line luminosity -- accretion relationship
has been derived empirically, using objects with \Lacc\ measured
independently from veiling or (for very low mass objects) from the
\Ha\ profiles, not from models.
The results of this  paper,
which suggests that  the IR lines may not necessarily
form in the accreting columns, do not weaken their reliability as a
tool to measure the accretion rate. 
In fact, if they  form in the accretion spots, they may provide a
measurement of \Macc\ in BDs in analogy to continuum veiling in TTS,
less model-dependent than the \Ha\ profiles used so far.

\appendix
\section {Spectral classification of the Ophiuchus objects}

\begin{figure}[ht!]
	\begin{center}
  \psfig{figure=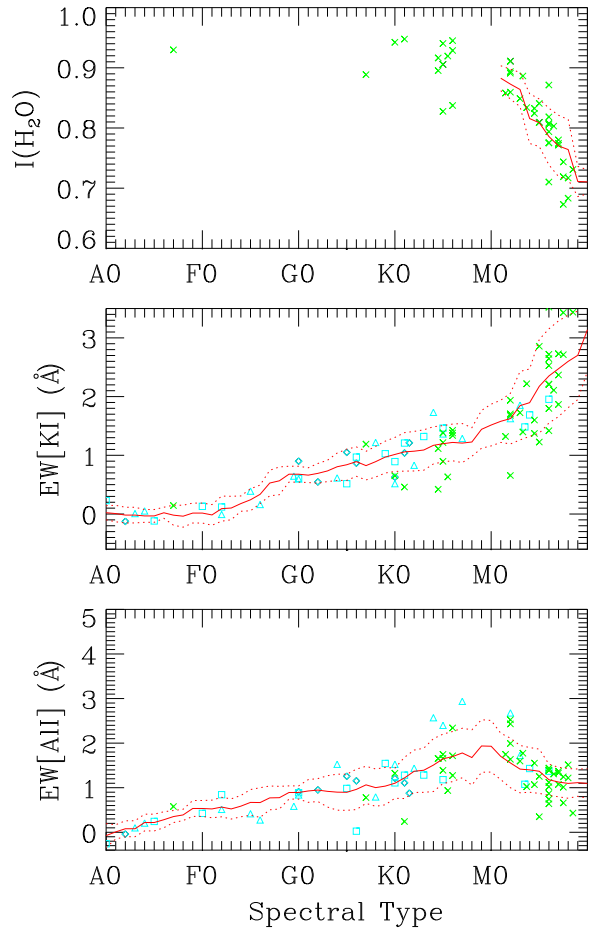, width=10cm,angle=0} 
\linespread{1.0}
\caption{
Run of the main spectral indices used for the spectral classification as a 
function of the spectral type. The top panel shows an index that measures the 
water vapour absorption at the red end of the J-band, the middle and bottom 
panels show the equivalent width of the [KI] and [AlII] lines within the 
J-band. Open squares, triangles and diamonds show the values of the [KI] and 
[AlII] equivalent widths computed for the giants (luminosity class III), dwarfs 
(V) and subgiants (IV) of Wallace et al.~(2000), respectively. Crosses show the
values of the same indices for the objects with known spectral type in Ophiuchus. 
The solid lines are smoothed averages of the measured points, the dotted lines
show the dispersion within the smoothing interval ($\pm 3$ subclasses); they 
have been used to estimate the spectral type range
for our target stars from the measured indices.}
\label{j_index}
\end{center}
\end {figure}

Many of our targets do not have a reliable spectral classification in the literature, others have been classified using a variety of different methods, from low resolution optical, to K-band or high resolution infrared spectroscopy. We attempted to use our J-band spectra to estimate in a uniform fashion the spectral types of all objects in our sample. 
The J-band was chosen because at this wavelength the photosphere dominates the emission even for Class II objects and
the effect of veiling from the disk infrared excess emission is thus limited,
especially in the BDs.

We computed the equivalent widths for a number of photospheric lines, following Wallace~(\cite{Wea00}). Due to the lower resolution of our spectra, we needed to adjust the limits for the computation 
of the equivalent widths given in Table~3 of Wallace et al.; thus, we
recomputed the values of the equivalent widths for all spectra in their list. We also computed the equivalent width values for all 
$\rho$-Oph Class~II objects in our Natta et al.~(\cite{Nea06}) sample with a solid spectral classification in the literature, mainly from Luhman \& Rieke~\cite{LR99} or from Natta et al.~\cite{Nea02}. To improve
the possibility of a good spectral classification at late types we also defined a spectral index based on the water absorption features at the red edge of the band, in a similar fashion as Testi et al.~(\cite{Tea01}). This index was computed only for the $\rho$-Oph Class~II objects with known spectral type. In Figure~\ref{j_index}, we show the run of the equivalent widths of [AlII] and [KI] and the water index as a function of the known spectral type. For late spectral types (K-M) a combination of the various indices may allow a reasonable classification within a few subclasses, at earlier spectral types the correlations of the indices with the spectral type flattens significantly and the classification is more uncertain (if at all possible). 

Using the equivalent widths and the water index, for each of the stars in our sample we derived a range of possible spectral types compatible with the measured values. These are reported in Table~\ref{ST} along with the spectral types found in the literature. In general our estimate of the spectral type is consistent with the values found in the literature. Following this comparison, for each target, we adopt a fiducial spectral type that is used throughout this paper to derive the stellar parameters (Column 9). For the very low
mass stars and brown dwarfs studied in Natta et al.~(\cite{Nea02}) we adopt their spectral type as
our spectral type range is essentially based on the water index and is more uncertain than the classification made using the complete near infrared spectrum. 
For these objects the adopted type is always within the range we derive here, with the exception of $\rho$Oph-ISO~193, for which an earlier type (by 2 subclasses)
is possible. For the stellar objects, the adopted spectral type is at the center of the range derived here, and it is usually consistent with the estimates in the literature. Some of the objects, however, have very uncertain spectral classification, this is reflected in the larger uncertainties in the effective temperature estimate used in Sect.~\ref{s2_1} and Fig.~\ref{fig-HR}.

\begin{landscape}
\tiny{
\begin{table}
\begin{center}
\caption {  Line indexes and derived spectral types. 
Column 1: name of object; 2, 3, 4: H$_2$O, KI and AlI indexes measured
from this paper spectra; 5, 6, 7: same from Natta et al.~(\cite{Nea06}) J-band spectra;
column 8: spectral type range; column 9: adopted spectral type;
column 10--13: spectral types from Natta et al.~(\cite{Nea02}), Luhman and Rieke (\cite{LR99}), Wilking et al.~(\cite{Wea05}) and Doppmann et al.~(\cite{Dea05}), respectively; column 14: other source names.}
\label{ST}
\begin{tabular}{lccccccccccccc}
Name & H$_2$O& K I& Al I& H$_2$O & K I& Al I& ST&ST& ST& ST&  ST&ST&Other Names\\
& \multicolumn{3}{c}{This Paper} & \multicolumn{3}{c}{Nea06} & 
range&adopted&  Nea02& LR99& Wea05& Dea05 &  \\
\hline 
\ISO 002a& 0.95& 1.9& 2.1& 0.88& 1.8& 1.6& K8--M2&M0& --& --& --&--&B162538-242238\\
\ISO 023&  0.75& 1.5& 0.5& 0.78& 2.7& 1.1&M4--M9&  M7& M7& --&--&--& SKS1\\
\ISO 030& 0.79& 2.2& 1.2& 0.82& 2.2& 1.4  & M4-M8&M6&M6& M5-M6&M5.5&--& GY5\\
\ISO 033&  0.71& 2.7& 0.7& 0.73& 3.4& 0.5&M7--M9&  M8.5& M8.5&--& --&--&GY11\\ 
\ISO 037 & 0.99& 1.0& 1.9& 0.92& 1.7& 2.5&K0--K9&  K5& --& K-M& --&K9& LFAM3/GY21\\
\ISO 083 & 0.93& 1.9& 2.0& 0.92& 1.7& 1.4&K8--K9& K9& --& --&--& --&BL162656-241353\\
\ISO 102&  0.81& 2.1& 1.1& 0.80& 2.7& 1.1&M3--M6&  M6& M6& --&M5.5&--& GY204\\
\ISO 105 & 0.93& 1.5& 1.6& 0.92& 1.1& 1.8&K8--M0& K9& --& K5-M2&--& M3& WL17/GY205\\
\ISO 115 & 0.85& 0.8& 1.2& 0.88& 1.0& 2.1&K9--M2&M0  &--& --& --&--&WL11/GY229\\
\ISO 117 &0.93& 1.4& 1.3&  0.86& 1.2& 2.4&K2--M3& K8& --& --& --&--&GY235\\
\ISO 155 &0.96& 0.9 & 1.4&  0.94& 0.8& 1.3 & K0--K5& K3& -- & K5-M2&--&--& GY292\\
\ISO 160&  0.81& 2.2& 1.2& 0.81& 2.6& 1.4&M3--M8 & M6& M6& --&--&--& B162737-241756\\
\ISO 163& 0.96& 1.2& 1.6& 0.91& 1.3& 1.8&K0--K8& K5& --& K5-M2& --&--&IRS49/GY308\\
\ISO 164 & 0.70& 3.0& 1.0& 0.71& 3.5& 0.7 &M7--M9 &M6&M6& --& M3-M5&--& GY310\\
\ISO 166 & 0.93& 1.6& 2.1& 0.91& 1.4& 1.8&K2--K8& K5& --& K6-M2& K0&--&GY314\\
\ISO 193&  0.86& 1.4& 1.3& 0.87& 1.4& 0.8&M1--M4& M6&  M6&--& --& --&B162812-241138\\
\hline
\end{tabular}
\end{center}
\end{table}
}
\end{landscape}

{}

\end{document}